\documentclass[journal]{IEEEtran}
\usepackage{latexsym}
\usepackage{graphicx}
\usepackage{amsfonts,amssymb,amsmath}
\usepackage{ctable} 
\usepackage{mathtools, cuted}
\usepackage{hyperref}
\usepackage[ruled,linesnumbered,noend]{algorithm2e}
\usepackage[T1]{fontenc}
\usepackage{cite}
\usepackage{subcaption}
\usepackage{comment}
\usepackage{diagbox}

\usepackage{amsthm}
\usepackage{overpic}
\usepackage{steinmetz}
\usepackage{array}
\usepackage{url}
\usepackage{color}

\usepackage{algorithmicx}
\usepackage{algcompatible}

\usepackage{xcolor}
\usepackage{soul}

\usepackage{linegoal}

\theoremstyle{plain}

\newcommand{\argmax}[1]{{\underset{{#1}}{\mathrm{arg\,max}}}}
\newcommand{\argmin}[1]{{\underset{{#1}}{\mathrm{arg\,min}}}}

\def\diag{\mathrm{diag}}

\def\Htran{\mbox{\tiny $\mathrm{H}$}}
\def\Ttran{\mbox{\tiny $\mathrm{T}$}}
\SetCommentSty{mycommfont}

\begin{document}
\bstctlcite{BSTcontrol}
\title{Maximum A Posteriori Probability Channel Tracking with an Intelligent Transmitting Surface}

\author{\IEEEauthorblockN{\normalsize Parisa Ramezani, \textit{Member, IEEE},
    Alva Kosasih, \textit{Member, IEEE}, and Emil Bj\"ornson, \textit{Fellow, IEEE}, }
\thanks{P. Ramezani and E. Björnson are with the KTH Royal Institute of Technology, Stockholm, Sweden. A. Kosasih is with Nokia Standards, Espoo, Finland. This work was supported by Digital Futures and the Swedish Innovation Agency (Vinnova) through the SweWIN center (2023-00572).}
}

\maketitle

\begin{abstract}
This paper considers an intelligent transmitting surface (ITS) integrated into a base station and develops a low-overhead maximum a posteriori (MAP) probability channel tracking method for the dominant line-of-sight link between the ITS and the user equipment. We cast the per-block channel as a three-parameter model consisting of the channel amplitude, channel phase, and angle-of-arrival at the ITS. We exploit temporal correlation by updating the priors using the estimates from the previous block. Using only two pilots per coherence block alongside a targeted beam alignment strategy, the proposed method achieves precise channel tracking and attains spectral efficiency close to that achievable under perfect channel knowledge.
\end{abstract}
\begin{IEEEkeywords}
Intelligent transmitting surface, maximum a posteriori probability, channel tracking.
\end{IEEEkeywords}

\vspace{-2mm}

\section{Introduction}
Owing to its unique ability to create virtual line-of-sight (LoS) links and enhance the rank of the channels, reconfigurable intelligent surface (RIS) has lately attracted extensive research interest, evolving from a purely theoretical concept to early hardware prototypes and experimental demonstrations \cite{Pei2021RIS,Chen2021Design,Zeng2024RIS,Li2024User,Gholami2025Wireless}. 
While the term RIS is often used to encompass both reflecting- and transmitting-type (meta)surfaces, the vast majority of existing works focus on reflecting-type RISs.

In the transmissive type of surface, which we refer to as an intelligent transmitting surface (ITS) throughout this paper, a few active antennas illuminate a large array of passive elements that re-transmit phase-shifted versions of the incoming signals. In \cite{Jamali2021Intelligent}, ITS-aided transmitter architectures are proposed for millimeter-wave (mmWave) systems, showing that a few radio frequency (RF) chains can efficiently feed an ultra-massive passive aperture and achieve highly energy-efficient precoding. Cell-free networks with ITS-integrated access points are studied in \cite{Demir2024User}, where ITS and active antennas are placed at opposite sides of an enclosed box with perfectly reflecting walls on the sides. A related line of work is \cite{Huang2024Hybrid}, where a single feed antenna illuminates an ITS whose elements can operate either in an active (amplifying) or passive mode.

A major challenge in RIS-assisted networks, whether reflective or transmissive, is the acquisition of accurate channel state information (CSI) due to the existence of many passive elements. The CSI is crucial to realizing the potential performance gains. Most existing channel estimation schemes have been developed targeting reflecting-based RISs, yet the underlying training overhead issues are equally present in ITS-based architectures.
Reference \cite{Emil2022Maximum} adopts a parametric maximum-likelihood (ML) approach, where instead of estimating one complex coefficient per RIS element, the channel is assumed to lie in a one-dimensional manifold defined by the RIS array response. The authors then perform a one-dimensional search over the angle-of-arrival (AoA), and compute the corresponding gain and phase in closed form. By adaptively updating the RIS configurations based on the current ML estimate, they estimate the channel with far fewer pilot transmissions than the number of RIS elements.
In \cite{Wu2023Parametric}, an RIS-assisted terahertz system is considered with a parametric sparse channel model, where the channel estimation problem is formulated as a structured sparse recovery problem, enabling path-parameter estimation with substantially reduced pilot overhead compared to non-parametric channel estimation.
Reference \cite{Li2024Channel} builds a covariance-based model for the channel between the RIS and user equipments (UEs), and estimates the channels via an eigenspace projection algorithm that projects the channels onto a low-dimensional eigenspace.  
The authors in \cite{Li2024Low} represent the RIS-cascaded channels in a lower-dimensional subspace via a Karhunen–Loève transform, and estimate only the dominant spatial eigenmodes using a linear minimum mean square error estimator. The required number of pilots in \cite{Li2024Channel} and \cite{Li2024Low} depends on the number of retained eigenmodes, with more modes requiring more pilots but yielding higher accuracy.
A low-dimensional orthogonal codebook of RIS configurations is developed in \cite{Haghshenas2024Parametric}, which is employed in an adaptive RIS configuration strategy that first performs a widebeam initialization using a couple of broad beams to roughly localize the UE. The RIS beams are then iteratively refined within this codebook to steer energy toward the UE and update the parametric channel estimates using only a few pilot transmissions. 

If prior knowledge about the channel is available, the required pilot overhead can be significantly reduced by exploiting it in the estimation procedure. Building on this concept, reference \cite{Zhou2022Channel} first estimates all channel parameters, including angles and channel gains in the first coherence block (CB). Then, the authors assume that the estimated angles remain constant over subsequent CBs and only re-estimate the path gains for later blocks. However, when the angles change, the scheme starts over again and forgets previous estimates, so there is no attempt to track the UE under large-scale mobility.
As the RIS configuration mainly depends on the angle, reliable UE angle tracking is essential in maintaining a stable, high-quality connection over time using minimal pilot resources. 

In this paper, we consider an ITS-integrated base station (BS) and propose a maximum a posteriori (MAP) probability channel tracking scheme that leverages prior knowledge of the UE-ITS channel evolution across CBs. Specifically, we model the LoS UE-ITS link  by three parameters, namely channel amplitude, channel phase, and AoA at the ITS, and track their inter-CB evolution using priors centered at the previous estimates.  This yields a MAP problem solved iteratively based on closed-form solutions for the channel amplitude and phase, and a grid search for the AoA. Each CB uses only two pilots with ITS configurations chosen as slight perturbations around the previously spectral efficiency (SE)-maximizing beam to probe the angular neighborhood of the current UE direction. We demonstrate via numerical simulations that, with only two pilots per CB, the proposed scheme accurately follows the channel evolution and outperforms a purely ML-based estimator that does not exploit prior information. 

\section{System Model}
Consider a BS architecture consisting of $N$ active antennas, each equipped with its own dedicated RF chain. An ITS  with $M$ elements is integrated into the antenna array, re-radiating a phase-shifted version of the impinging signal. We consider uplink pilot transmission, where the UE sends pilots to the BS for channel estimation and tracking.

The channel between the active antennas and the ITS consists of a LoS component and a non-LoS (NLoS) component. We denote by $\mathbf{H}_{\mathrm{LoS}} \in \mathbb{C}^{M \times N}$ the LoS part of the channel, which can be expressed as 
\begin{align}
  [\mathbf{H}_{\mathrm{LoS}}]_{m,n} &= \rho_0 \sqrt{G_{1}(\bar{\phi}_{m,n},\bar{\psi}_{m,n}) G_{2}(\tilde{\phi}_{m,n},\tilde{\psi}_{m,n})} \nonumber \\
  &\quad \cdot \frac{\lambda}{4\pi d_{m,n}} e^{-j\frac{2\pi}{\lambda} d_{m,n}},
\end{align}
where $G_{1}(\cdot)$ and $G_{2}(\cdot)$ respectively represent the directional patterns of ITS elements and active antennas, $(\bar{\phi}_{m,n},\bar{\psi}_{m,n})$ is the angle pair describing the direction under which the $n$-th antenna is seen from the $m$-th ITS element, and $(\tilde{\phi}_{m,n},\tilde{\psi}_{m,n})$ is the angle pair showing the direction under which the $m$-th ITS element is seen from the $n$-th antenna. $\lambda$ is the carrier wavelength and $d_{m,n}$ is the distance between the $m$-th ITS element and $n$-th antenna. 
Furthermore, $\rho_0$ is an attenuation factor that accounts for ohmic and dielectric losses in the ITS and enclosure, non-perfectly conducting walls, finite radiation efficiency of the active antennas and ITS elements, and possible mismatch or feeder losses.
The NLoS component $\mathbf{H}_{\mathrm{NLoS}} \in \mathbb{C}^{M \times N}$ is given by 
\begin{equation}
  \mathbf{H}_{\mathrm{NLoS}} = \sqrt{\rho_{\mathrm{NLoS}}}\, \mathbf{R}_{\mathrm{ITS}}^{1/2}\mathbf{W}\mathbf{R}_{\mathrm{Ant}}^{1/2},  
\end{equation}
where $\mathbf{W} \in \mathbb{C}^{M \times N}$ entries are i.i.d. $\mathcal{CN}(0,1)$-distributed, $\mathbf{R}_{\mathrm{ITS}}$ and $\mathbf{R}_{\mathrm{Ant}}$ are the correlation matrices on the ITS and active antennas sides, respectively. We assume that the NLoS component is $10\,$dB weaker than the LoS component. So, $\rho_{\mathrm{NLoS}} = P_{\mathrm{LoS}}/10$, where $P_{\mathrm{LoS}} = \frac{1}{MN}\sum_{m=1}^M \sum_{n=1}^N \left|[\mathbf{H}_{\mathrm{LoS}}]_{m,n}\right|^2$. The overall channel between the active antennas and the ITS is $\mathbf{H} = \mathbf{H}_{\mathrm{LoS}} + \mathbf{H}_{\mathrm{NLoS}}$.

For the purpose of exposition and notational clarity, we assume one active antenna and a linear one-dimensional ITS in this paper. Hence, $\mathbf{H}$ reduces to a vector $\mathbf{h} \in \mathbb{C}^M$, assumed fixed and known. The generalization to multiple active antennas and a planar ITS is immediate and does not alter the subsequent analysis.

\subsection{MAP Channel Estimation Problem}
Our objective is to track a mobile UE by exploiting prior knowledge about the UE's movement and the channel parameters. We focus on the dominant LoS component of the UE-ITS link and model the UE-ITS channel as a rank-one LoS channel.
Specifically, the UE-ITS channel in the $t$-th CB is modeled as $\mathbf{g}_t = \beta_t e^{j\omega_t}\mathbf{a}(\varphi_t)$, where $\beta_t$ is the channel amplitude, $\omega_t$ is the channel phase,  and $\mathbf{a}(\varphi_t) \in \mathbb{C}^{M}$ denotes the array response vector with the AoA of $\varphi_t$. In the $t$-th CB, the UE transmits the deterministic pilot signal $x = \sqrt{P_{\mathrm{p}}}$ over $L_t$ time instants, while the ITS applies a different phase-shift configuration at each instant. Concatenating the received signals at the BS over $L_t$ time instants, we obtain
\begin{equation}
\label{eq:received_pilots}
    \mathbf{y}_t = \sqrt{P_{\mathrm{p}}}\boldsymbol{\Theta}_t\mathbf{D}_{\mathrm{h}} \mathbf{g}_t +  \mathbf{n}_t,
\end{equation}
where $\mathbf{y}_t = [y_{t,1},\ldots,y_{t,L_t}]^{\Ttran} \in \mathbb{C}^{L_t}$ with $y_{t,l}$ being the received signal at time instant $l$, and $\boldsymbol{\Theta}_t = [\boldsymbol{\theta}_{t,1},\ldots,\boldsymbol{\theta}_{t,L_t}]^{\Ttran}$ with $\boldsymbol{\theta}_{t,l}$ being the ITS configuration vector at time instant $l$. Further, $\mathbf{D}_{\mathrm{h}} = \diag(\mathbf{h})$, and $\mathbf{n}_t = [n_{t,1},\ldots,n_{t,L_t}]^{\Ttran}$ denotes the noise at all pilot transmission instants in the $t$-th CB. 

For each CB, we estimate the UE-ITS channel using a MAP approach. Let $\mathbf{Y}_t$ denote the random received pilot vector and $\mathbf{y}_t$ its realization. We obtain $\hat{\mathbf{g}}_t$ as the maximizer of the posterior density of the channel as 
\begin{equation}
\hat{\mathbf{g}}_t = \arg\max_{\mathbf{g}_t} f_{\mathbf{Y}_t}(\mathbf{y}_t; \mathbf{g}_t) f(\mathbf{g}_t),
\end{equation}
with 
\begin{align}
  f_{\mathbf{Y}}(\mathbf{y}_t;\mathbf{g}_t) &= \frac{1}{(\pi \sigma^2)^{L_t}} e^{-\frac{\|\mathbf{y}_t - \sqrt{P_{\mathrm{p}}}\boldsymbol{\Theta}_t\mathbf{D}_{\mathrm{h}}\mathbf{g}_t\|^2}{\sigma^2}} \nonumber \\
  & = \frac{1}{(\pi \sigma^2)^{L_t}} e^{- \frac{\|\mathbf{y}_t - \sqrt{P_{\mathrm{p}}}\beta_t e^{j\omega_t}\boldsymbol{\Theta}_t\mathbf{D}_{\mathrm{h}}\mathbf{a}(\varphi_t)\|^2}{\sigma^2}},
\end{align}
and 
\begin{equation}
\label{eq:prior_distribution}
   f(\mathbf{g}_t) = f(\beta_t)f(\omega_t)f(\varphi_t),
\end{equation}
where $f_{\mathbf{Y}}(\mathbf{y}_t; \mathbf{g}_t)$ is the likelihood, i.e., the probability density function (PDF) of $\mathbf{Y}_t$ given $\mathbf{g}_t$ evaluated at $\mathbf{y}_t$, and $f(\mathbf{g}_t)$ is the prior PDF of $\mathbf{g}_t$. Further, $\sigma^2$ is the noise variance and \eqref{eq:prior_distribution} holds under the standard assumption that the channel amplitude, channel phase, and AoA are a priori independent. 

\vspace{-2mm}
\subsection{Prior Distributions of the Channel Parameters}
For $\beta_t$ and $\varphi_t$, we adopt Gaussian priors centered at the previous estimates, i.e., 
\begin{align}
\label{eq:dis_beta}f(\beta_t) &= \frac{1}{\sqrt{2\pi \sigma_{\beta}^2}}    e^{-\frac{(\beta_t-\mu_\beta)^2}{2\sigma_\beta^2}}, \\
f(\varphi_t)  &= \frac{1}{\sqrt{2\pi \sigma_{\varphi}^2}}    e^{-\frac{(\varphi_t-\mu_\varphi)^2}{2\sigma_\varphi^2}} \label{eq:dist_varphi},
\end{align}
where in the $t$-th CB, we set $\mu_\beta = \hat{\beta}_{t-1}$ and $\mu_{\varphi} = \hat{\varphi}_{t-1}$. 
These priors are motivated by the smooth evolution of both parameters across CBs, which we model as small Gaussian deviations from the previous estimates. By the central limit theorem, the aggregate effect of many small inter-CB perturbations is approximately Gaussian. The variances $\sigma_\beta^2$ and $\sigma_\varphi^2$ reflect the confidence level in the previous estimates. For the AoA, we assume $\varphi_t \in [\hat{\varphi}_{t-1} - \Delta_\varphi, \hat{\varphi}_{t-1} + \Delta_\varphi]$ with high probability. Here, $\Delta_\varphi$ represents the half-width of the angular uncertainty interval around the previous AoA estimate  $\hat{\varphi}_{t-1}$. 
Furthermore, we use the three-sigma rule to bound the parameter uncertainty, since about $99.7 \%$ of a Gaussian’s probability mass lies within three standard deviations. We thus set $\sigma_\varphi^2 = (\Delta_\varphi/3)^2$. 

The phase $\omega_t$ is a circular variable on $[0,2\pi)$, making an unbounded Gaussian prior unsuitable. We therefore model $\omega_t$ with a von Mises distribution, the circular analogue of the Gaussian that captures periodicity and phase wrap-around,
 i.e., 
\begin{equation}
 \label{eq:dist_omega} f(\omega_t) = \frac{1}{2\pi I_0(\kappa)} e^{\kappa \cos(\omega_t - \mu_\omega)},
\end{equation}
where $I_0(\kappa)$ is the modified Bessel function of order $0$. $\kappa$ is a shape parameter, called the ``concentration'' providing a quantitative measure of phase uncertainty, where a large $\kappa$ makes the distribution concentrated around $\mu_{\omega}$, while the distribution becomes uniform for $\kappa = 0$. We set $\mu_\omega = \hat{\omega}_{t-1}$ and adjust $\kappa$ based on how we expect the channel phase to vary from one CB to the next.

\section{MAP Channel Tracking}
\label{sec:MAP_tracking}
In this section, we aim to estimate the parameters $\beta_t,\,\omega_t,\,\varphi_t$ from the received pilot vector $\mathbf{y}_t$ in \eqref{eq:received_pilots}, while exploiting the prior information from the previous CB collected in \eqref{eq:dis_beta}-\eqref{eq:dist_omega}. The joint posterior PDF of $(\beta_t, \omega_t, \varphi_t)$, up to a normalization constant, is 
\begin{align}
f_{\beta_t,\omega_t,\varphi_t|\mathbf Y_t}&(\beta_t,\omega_t,\varphi_t \mid \mathbf y_t)
\propto \nonumber \\
\exp\Bigg(
&-\frac{\big\|\mathbf y_t - \sqrt{P_p}\,\beta_t e^{j\omega_t}\boldsymbol{\Theta}_t\mathbf D_{\mathrm{h}}\mathbf a(\varphi_t)\big\|^2}{\sigma^2}
\notag\\
&-\frac{(\beta_t-\mu_{\beta})^2}{2\sigma_\beta^2}
-\frac{(\varphi_t-\mu_{\varphi})^2}{2\sigma_\varphi^2}
+\kappa\cos(\omega_t-\mu_{\omega})
\Bigg).
\label{eq:posterior_explicit}
\end{align}
The MAP estimates are thus given by 
\begin{equation}
 \{\hat{\beta}_t,\hat{\omega}_t,\hat{\varphi}_t\} = \argmax{\beta_t \geq 0,\omega_t,\varphi_t \in \Phi_t}  f_{\beta_t,\omega_t,\varphi_t|\mathbf Y_t}(\beta_t,\omega_t,\varphi_t \mid \mathbf y_t), 
\end{equation}
where $\Phi_t = [\hat{\varphi}_{t-1}-\Delta_\varphi, \hat{\varphi}_{t-1} + \Delta_\varphi] \, \cap \, [-\pi/2,\pi/2]$ is the expected range for the AoA. 
Equivalently, we can minimize the negative log-posterior. Let $\mathbf{b}_t(\varphi_t) = \boldsymbol{\Theta}_t \mathbf{D}_{\mathrm{h}}\mathbf{a}(\varphi_t) \in \mathbb{C}^{L_t}$. By expanding the squared norm, we have
\begin{align}
   &\left\|\mathbf{y}_t - \sqrt{P_{\mathrm{p}}} \beta_t e^{j\omega_t} \mathbf{b}_t(\varphi_t)\right\|^2  = \|\mathbf{y}_t\|^2 + \nonumber\\
   & P_{\mathrm{p}} \beta_t^2 \|\mathbf{b}_t(\varphi_t)\|^2 - 2\sqrt{P_{\mathrm{p}}} \beta_t \Re\{e^{j\omega_t}\mathbf{y}_t^{\Htran} \mathbf{b}_t(\varphi_t)\}.
\end{align}
Dropping the constant term $\|\mathbf{y}_t\|^2$ and scaling the resulting negative log-posterior by $\sigma^2$, the MAP problem becomes
\begin{equation}
  \{\hat{\beta}_t,\hat{\omega}_t,\hat{\varphi}_t\} = \argmin{\beta_t \geq 0,\omega_t,\varphi_t \in \Phi_t} J_t(\beta_t,\omega_t,\varphi_t),   
\end{equation}
where
\begin{align}
 &J_t(\beta_t,\omega_t,\varphi_t) =    P_{\mathrm{p}} \beta_t^2 \|\mathbf{b}_t(\varphi_t)\|^2 - 2\sqrt{P_{\mathrm{p}}} \beta_t \Re\{e^{j\omega_t}\mathbf{y}_t^{\Htran} \mathbf{b}_t(\varphi_t)\} \nonumber \\
 &+ \gamma_\beta(\beta_t - \mu_\beta)^2+ \gamma_\varphi(\varphi_t - \mu_\varphi)^2 - \gamma_\omega \cos(\omega_t - \mu_\omega),
\end{align}
with 
\begin{equation}
 \gamma_\beta = \frac{\sigma^2}{2\sigma_\beta^2},\,\, \gamma_\varphi = \frac{\sigma^2}{2\sigma_\varphi^2},\,\, \gamma_\omega = \sigma^2 \kappa.    
\end{equation}
In the following, we derive the MAP estimates by optimizing the objective function with respect to each parameter, while keeping the other two fixed.  
\vspace{-3mm}
\subsection{Optimization with Respect to $\omega_t$}

For fixed $\beta_t$ and $\varphi_t$, only the second and the last term of $J_t(\beta_t,\omega_t,\varphi_t)$ depend on $\omega_t$. So, $\omega_t$ can be estimated as 
\begin{equation}
    \hat{\omega}_t = \argmax{\omega_t}\,2\sqrt{P_{\mathrm{p}}}\beta_t \Re\{e^{j\omega_t}\mathbf{y}_t^{\Htran}\mathbf{b}_t(\varphi_t)\} + \gamma_\omega \cos(\omega_t - \mu_\omega).
\end{equation}
Using $\cos(\omega_t - \mu_\omega) = \Re\{e^{j(\omega_t - \mu_\omega)}\}$, we obtain 
\begin{equation}
  \hat{\omega}_t = \argmax{\omega_t}\, \Re \left\{e^{j\omega_t}(2\sqrt{P_{\mathrm{p}}} \beta_t\mathbf{y}_t^{\Htran}\mathbf{b}_t(\varphi_t) + \gamma_\omega e^{-j\mu_\omega}) \right\},   
\end{equation}
which yields a closed-form solution for $\omega_t$ as 
\begin{equation}
 \hat{\omega}_t = -\arg(2\sqrt{P_{\mathrm{p}}} \beta_t\mathbf{y}_t^{\Htran}\mathbf{b}_t(\varphi_t) + \gamma_\omega e^{-j\mu_\omega}).    
\end{equation}

\subsection{Optimization with Respect to $\beta_t$}
For fixed $\omega_t$ and $\varphi_t$, $J_t(\beta_t,\omega_t,\varphi_t)$ is a quadratic function of $\beta_t$:
\begin{align}
 &J_t(\beta_t,\omega_t,\varphi_t) =  \nonumber \\&(P_{\mathrm{p}} \|\mathbf{b}_t(\varphi_t)\|^2 + \gamma_\beta)\beta_t^2 - 
 2(\sqrt{P_{\mathrm{p}}} \Re\{e^{j\omega_t}\mathbf{y}_t^{\Htran} \mathbf{b}_t(\varphi_t)\} + \gamma_\beta \mu_\beta)\beta_t  \nonumber \\
 &+\gamma_\beta \mu_\beta^2 + \gamma_\varphi(\varphi_t - \mu_\varphi)^2 - \gamma_\omega\cos(\omega_t - \mu_\omega).
\end{align}
The minimizer of $J_t(\cdot)$ with respect to $\beta_t$ is obtained by differentiating $J_t(\cdot)$ with respect to $\beta_t$ and setting the derivative equal to zero. Doing so, we obtain 
\begin{equation}
 \check{\beta}_t = \frac{\sqrt{P_{\mathrm{p}}} \Re\{e^{j\omega_t}\mathbf{y}_t^{\Htran} \mathbf{b}_t(\varphi_t)\} + \gamma_\beta \mu_\beta}{P_{\mathrm{p}} \|\mathbf{b}_t(\varphi_t)\|^2 + \gamma_\beta}.    
\end{equation}
Imposing the physical constraint $\beta_t \geq 0$, the MAP estimate of $\beta_t$ is 
\begin{equation}
  \hat{\beta}_t = \max(0,\check{\beta}_t). 
\end{equation}

\subsection{Optimization with Respect to $\varphi_t$}
The MAP estimate of $\varphi_t$, given $\omega_t$ and $\beta_t$, is obtained as 
\begin{align}
 \hat{\varphi}_t &= \argmax{\varphi_t \in \Phi_t}\, P_{\mathrm{p}} \beta_t^2 \|\mathbf{b}_t(\varphi_t)\|^2 \nonumber \\ &- 2\sqrt{P_{\mathrm{p}}} \beta_t \Re\{e^{j\omega_t}\mathbf{y}_t^{\Htran} \mathbf{b}_t(\varphi_t)\} + \gamma_{\varphi} \varphi_t^2 - 2\gamma_\varphi \varphi_t,   
\end{align}
where $\hat{\varphi}_t$ is obtained by performing a one-dimensional search over the angular interval $\Phi_t$.

\subsection{ITS Configuration for Pilot Transmission}
\begin{figure*}[!t]
  \centering

  \begin{subfigure}[t]{0.3\textwidth}
    \centering
    \includegraphics[width=\linewidth]{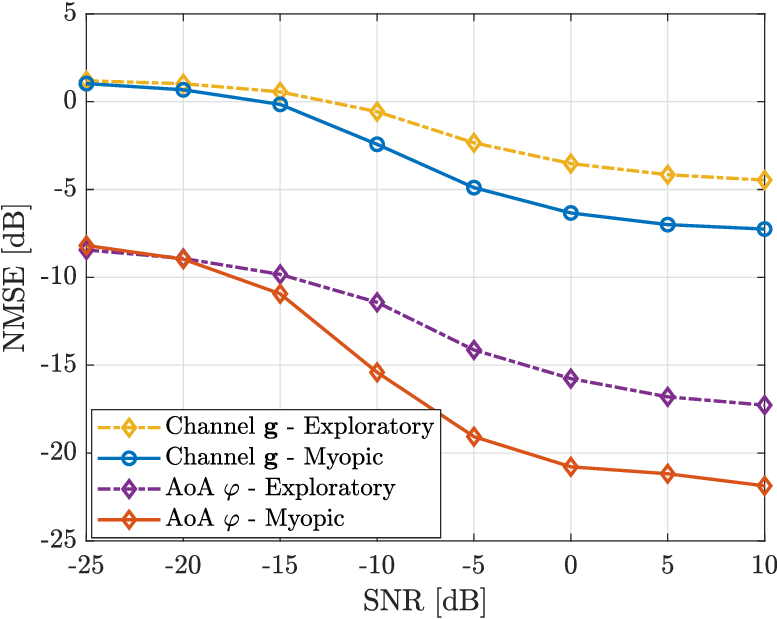}
    \caption{Comparison between myopic and exploratory channel tracking: conservative. }
    \label{fig:nmse1}
  \end{subfigure}\hfill
  \begin{subfigure}[t]{0.3\textwidth}
    \centering
    \includegraphics[width=\linewidth]{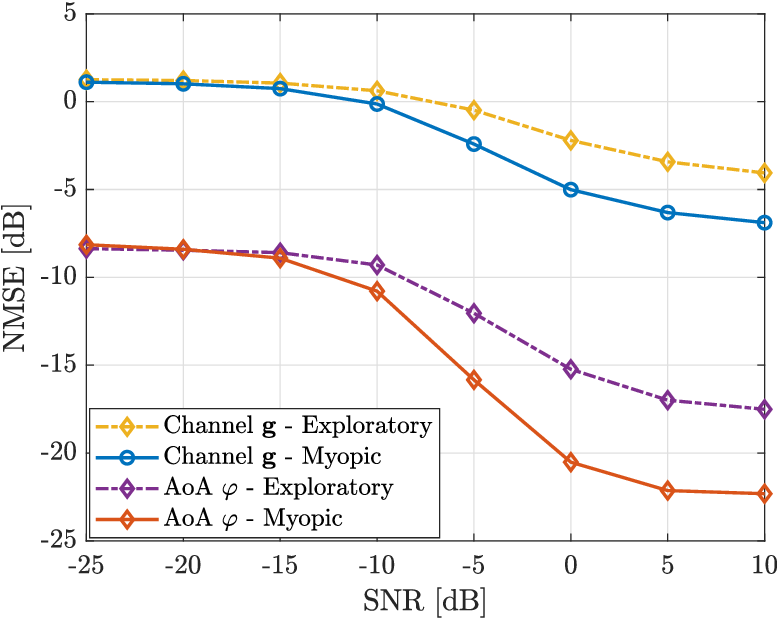}
    \caption{Comparison between myopic and exploratory channel tracking: over-confident.}
    \label{fig:nmse2}
  \end{subfigure}\hfill
  \begin{subfigure}[t]{0.3\textwidth}
    \centering
    \includegraphics[width=\linewidth]{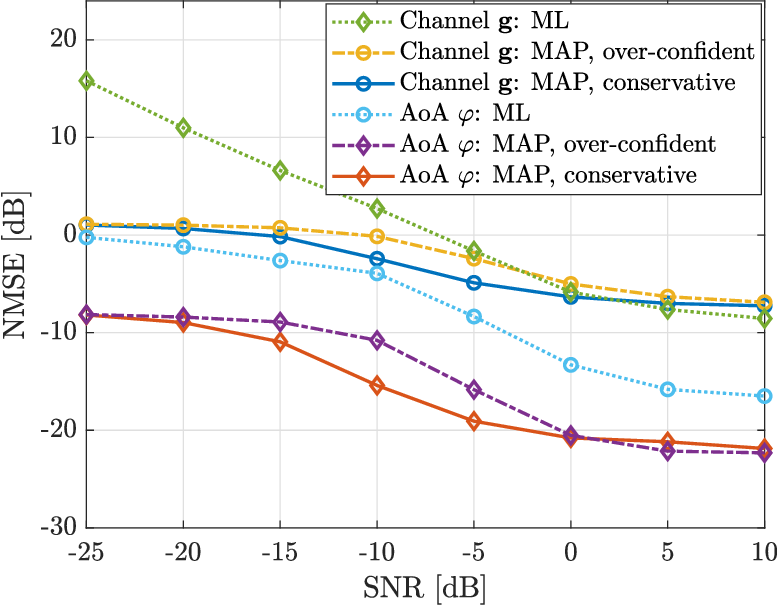}
    \caption{Comparison between MAP and ML channel estimation schemes.  }
    \label{fig:nmse3}
  \end{subfigure}

  \caption{NMSE versus SNR.}
  \label{fig:nmse_snr}
  \vspace{-2mm}
\end{figure*}
In order to track the UE with a very small pilot overhead, we restrict the number of pilot transmissions per CB to $L_t = 2$. With the proposed MAP estimator, this will be sufficient to track the channel variations. Here, we describe how the ITS is configured during pilot transmission to enable reliable MAP-based tracking of the UE-ITS channel.

Let $\bar{\boldsymbol{\theta}}_{t-1}$ be the SE-maximizing ITS configuration computed based on the estimated channel in the $(t-1)$-th CB. Specifically, having $\hat{\mathbf{g}}_{t-1}$ as the estimated channel in the $(t-1)$-th CB, the ITS configuration for maximizing the SE will be obtained as $\bar{\boldsymbol{\theta}}_{t-1} = e^{-j\arg(\mathbf{h} \odot \hat{\mathbf{g}}_{t-1})}$ \cite{Emil2022Maximum}, where $\odot $ denotes the element-wise multiplication.
Assuming a discrete codebook $C_\theta$ for candidate ITS configurations, we pick two RIS configurations $\boldsymbol{\theta}_{t,1},\,\boldsymbol{\theta}_{t,2} \in C_\theta$ to be used in the two pilot transmission instants in the $t$-th CB using a myopic tracking around the previous optimized angle.
In particular, we use 
\begin{equation}
    \label{eq:first_pilot}\boldsymbol\theta_{t,1}
    = \argmax{\boldsymbol\theta\in C_\theta}
    \big|\bar{\boldsymbol\theta}_{t-1}^{\Htran} \boldsymbol\theta\big|,
\end{equation}
\begin{equation}
\label{eq:second_pilot}
    \boldsymbol\theta_{t,2}
    = \argmax{\boldsymbol\theta\in C_\theta \setminus\{\boldsymbol\theta_{t,1}\}}
    \big|\bar{\boldsymbol\theta}_{t-1}^{\Htran} \boldsymbol\theta\big|,
\end{equation}
where \eqref{eq:first_pilot} selects $\boldsymbol{\theta}_{t,1}$ as the codeword whose beam pattern best aligns with the previously SE-maximizing configuration, and \eqref{eq:second_pilot} then chooses $\boldsymbol{\theta}_{t,2}$ as the next best-aligned codeword, so that the two configurations both steer beams close to the previous optimum while introducing a small local perturbation.

\section{Numerical Results}
In this section, we present simulation results to validate the effectiveness of the proposed MAP tracking scheme. Unless otherwise specified, the following parameters are used: we consider $M = 64$ ITS elements with half-wavelength spacing at the carrier frequency of $30\,$GHz, located along the $y-$axis, where the position of the $m$-th element is $(0,y_m,0)$ with $y_m = (m - \frac{M-1}{2})\frac{\lambda}{2}$. The active antenna is placed at $(-1,0,0)$. We use a cosine directional pattern for the ITS elements and the active antenna, given by $G_1(\bar{\phi}_m) = 2\cos^2(\bar{\phi}_m),\,G_2(\tilde{\phi}_m) = 2\cos^2(\tilde{\phi}_m)$, where $\cos(\bar{\phi}_m) = \cos(\tilde{\phi}_m) = \frac{1}{\sqrt{1+y_m^2}}$.  We set $\rho_0 = 1$ without loss of generality.
We evaluate the performance after $50$ CBs. Specifically, assuming an accurate initial estimate $(\hat{\beta}_0,\hat{\omega}_0,\hat{\varphi}_0)$, which can be obtained using the ML approach in \cite{Emil2022Maximum}, we run the proposed MAP channel tracking over $50$ blocks and report the results at the end of the $50$th block. The initial channel phase $\hat{\omega}_0$ and AoA $\hat{\varphi}_0$ are taken randomly from $\mathrm{Uniform}[-\pi,\pi]$ and $\mathrm{Uniform}[-\pi/4,\pi/4]$, respectively, and the initial channel amplitude is set as $\hat{\beta}_0 = 5\cdot 10^{-5}$.
In each CB, the channel parameters follow a first-order Markov model, according to  $\varphi_t \sim \mathcal{N}(\varphi_{t-1},\sigma_\varphi ^2)$, $\beta_t \sim \mathcal{N}(\beta_{t-1},\sigma_\beta^2)$, and $\omega_t \sim \mathrm{vonMises}(\omega_{t-1},\kappa)$ with $\sigma_\varphi = \pi/360$, $\sigma_\beta = 10^{-6}$, and $\kappa = 100$. $\beta_t$ is truncated to be non-negative. In the MAP estimation, we deliberately use mismatched hyperparameters $\sigma_{\varphi,\mathrm{est}} = f_\varphi \sigma_\varphi$, $\sigma_{\beta,\mathrm{est}} = f_\beta \sigma_\beta$, and $\kappa_{\mathrm{est}} = f_\kappa \kappa$, where $f_x$ denotes the random scaling factor that corresponds to parameter $x$. The AoA search interval will then be $\Phi_t = [\hat{\varphi}_{t-1} - 3\sigma_{\varphi,\mathrm{est}}, \hat{\varphi}_{t-1} + 3\sigma_{\varphi,\mathrm{est}}]$.
The DFT codebook is used for ITS configuration during pilot transmissions. 

To evaluate the performance of the proposed MAP tracking, we compare it against the following two benchmark schemes:

\begin{enumerate}
    \item \textbf{Exploratory MAP tracking:} In this scheme, we implement the same MAP estimator as in Section~\ref{sec:MAP_tracking}, but select the two ITS configurations in a more exploratory manner than the proposed scheme by using one deterministic beam and one randomized beam. The deterministic beam is designed by setting the ITS configuration as in \eqref{eq:first_pilot}. The second pilot uses a randomized configuration $\boldsymbol{\theta}_{t,2}$, drawn uniformly from the DFT codebook subset whose main beams point to angles within $\Phi_t$. 

    \item \textbf{ML estimation:} In this scheme, we apply an ML estimator and perform channel estimation in each CB without exploiting any prior information from previous blocks \cite{Emil2022Maximum}. We keep the pilot overhead identical to the proposed approach by using $L_t = 2$
    pilots per CB. 
   The corresponding ITS configurations for these pilots are selected according to the myopic design in \eqref{eq:first_pilot}--\eqref{eq:second_pilot}.
\end{enumerate}

We study two model-mismatch regimes between the true channel parameter and the priors assumed by the estimator, termed ``conservative'' and ``over-confident''. In the conservative regime, the estimator searches over an uncertainty region larger than the true one by drawing the scaling factors
$f_\varphi$ and $f_\beta$ i.i.d. from $\mathrm{Uniform}[1,2]$ and $f_\kappa$ from $\mathrm{Uniform}[0.5,1]$. In the over-confident scenario, we use a search region narrower than the true one by drawing $f_\varphi$ and $f_\beta$ i.i.d. from $\mathrm{Uniform}[0.5,1]$ and $f_\kappa$ from $\mathrm{Uniform}[1,2]$. Whenever the model-mismatch regime is not explicitly specified, the conservative case is assumed. 

Fig.~\ref{fig:nmse_snr} shows the normalized mean square error (NMSE) of the overall channel estimate and the AoA estimate as a function of signal-to-noise ratio (SNR), where the results are averaged over $1000$ Monte Carlo simulations. 
Fig.~\ref{fig:nmse1} and Fig.~\ref{fig:nmse2} compare the NMSE of myopic and exploratory designs for conservative and over-confident cases, respectively. We can see that the myopic design consistently outperforms the exploratory design across the SNR range. It is because with only two pilots, it is more effective to spend both measurements on high-gain, directionally consistent probing around the previous SE-maximizing codeword, rather than assign the second pilot to a randomly chosen beam that can be weakly aligned. The gap between the two schemes increases with the SNR since at high SNR, the estimation error is dominated by beam misalignment, which the myopic scheme mitigates more effectively. 
Fig.~\ref{fig:nmse3} shows that the proposed MAP tracking scheme (in both conservative and over-confident modes) outperforms ML in most operating regimes. ML becomes competitive at high SNR for the overall channel NMSE only, whereas MAP provides lower error at low-to-medium SNRs for the channel and consistently outperforms ML in  AoA estimation. 

\begin{figure}
    \centering
    \includegraphics[width=0.67\linewidth]{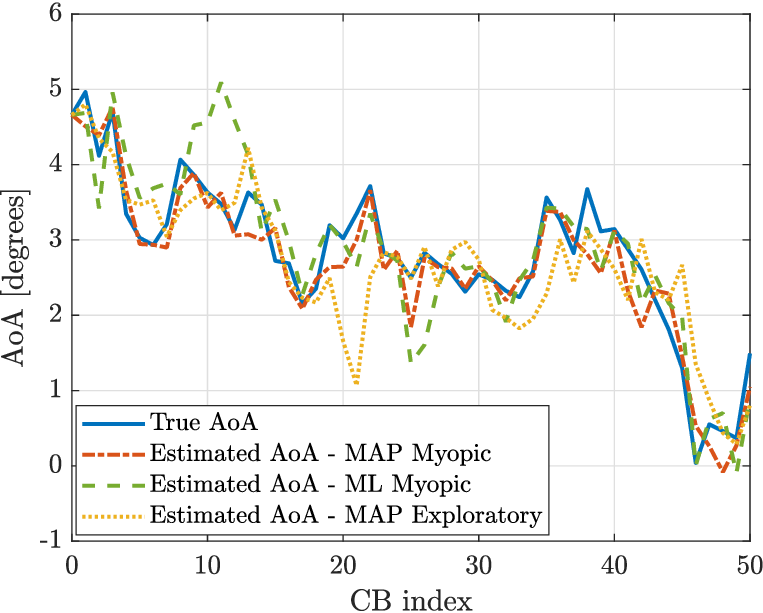}
    \caption{AoA tracking. }
\label{fig:AoA_tracking}
\vspace{-3mm}
\end{figure}
\begin{figure}
    \centering
    \includegraphics[width=0.67\linewidth]{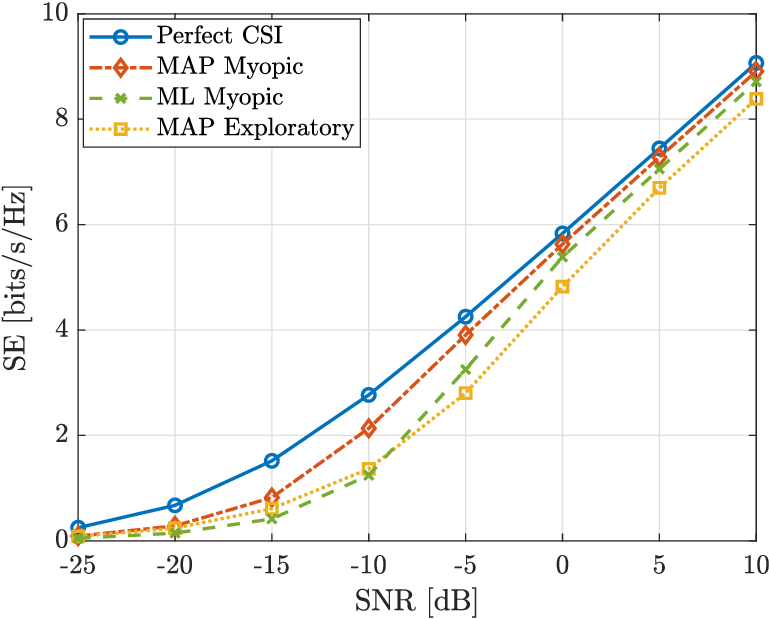}
    \caption{SE versus SNR. }
\label{fig:SE}
\vspace{-4.2mm}
\end{figure}
Fig.~\ref{fig:AoA_tracking} illustrates the AoA trajectory over $50$ CBs when SNR equals $5\,$dB. We can see that the proposed MAP-myopic scheme follows the true AoA most closely and exhibits the smallest deviations over time. In contrast, the MAP-exploratory and ML benchmarks show larger fluctuations and occasional noticeable tracking errors.

Finally, Fig.~\ref{fig:SE} shows that the proposed MAP-myopic scheme achieves an average SE that closely approaches the perfect CSI scenario in which the ITS is configured using the exact channel information. This indicates that the channel estimated by the proposed scheme is sufficiently accurate with only $L_t = 2$ pilots. Another observation is that the gap between the proposed scheme and ML is small at low SNR because both methods are primarily noise-limited. As SNR increases, MAP benefits more from exploiting temporal priors which translates into a larger SE advantage. At high SNR, however, ML's estimates become more accurate, which narrows its performance gap with the proposed MAP tracking. 
\vspace{-2mm}
\section{Conclusions}
In this paper, we introduced a MAP channel tracking scheme for systems employing ITS as part of the BS. By exploiting prior knowledge of channel parameter evolution across CBs, the proposed method estimates the dominant LoS ITS-UE link using only two pilots per block, while updating the ITS configuration via a myopic beam refinement strategy. The numerical results demonstrate that the proposed MAP tracking, together with the myopic beam refinement, provides accurate channel and AoA tracking, yielding effective performance in terms of both NMSE and SE. We note that the same MAP-based tracking principle can also be applied to reflective RIS deployments. 
\vspace{-2mm}
\bibliographystyle{IEEEtran}
\bibliography{refs} 
\end{document}